\documentclass{acm_proc_article-sp}
\usepackage{mathrsfs}
\usepackage[hidelinks]{hyperref}
\begin{document}

\title{Trust-Aware Neighbourhood model for Trust-based Recommendation}

\numberofauthors{4} 
\author{
\alignauthor
Amira Ghenai
\\ \affaddr{University of Waterloo, Mathematics Department}\\
       \affaddr{200 University Ave.W, Waterloo, ON N2L 3G5}\\
       \email{aghenai@uwaterloo.ca}
\alignauthor
Moustafa M. Ghanem\\
       \affaddr{Middlesex University, School of Science and Technology}\\
       \affaddr{The Burroughs, London NW4 4BT,  UK}\\
       \email{m.ghanem@mdx.ac.uk}
}

\date{11 February 2016}

\maketitle
\begin{abstract}
Traditional Recommender Systems (RS) do not consider any personal user information beyond rating history. Such information, on the other hand, is widely available on social networking sites (Facebook, Twitter). As a result, social networks have recently been used in recommendation systems. In this paper, we propose an efficient method for incorporating social signals into the recommendation process by building a trust network which supplements the users' rating profiles.
We first show the effect of different cold-start users types on the Collaborative Filtering (CF) technique in several real-world datasets. 
Inspired by \cite{Massa}, we propose a \emph{Trust-Aware Neighbourhood} algorithm which addresses a performance issue of the former by limiting the trusted neighbourhood. We show the doubling of the rating coverage compared to the traditional CF technique, and a significant improvement in the accuracy for some datasets. 
Focusing specifically on cold-start users, we propose a \emph{Hybrid Trust-Aware Neighbourhood} algorithm which expands the neighbourhood by considering both trust and rating history of the users. We show a near complete coverage with a rich trust network dataset-- Flixster. 
We conclude by discussing the potential implementation of this algorithm in a budget-constrained cloud environment.
\end{abstract}

\category{H.3.3}{Information Storage and Retrieval}{ Information Search and Retrieval}-- \emph{Information filtering}.


\keywords{Recommender Systems, Collaborative Filtering, Trust-Networks} 

\section{Introduction}
\setlength{\parindent}{0.3cm} 
\textnormal{\indent{Choosing} the right resource to get a recommendation on items is a critical component in human decision making. With the emergence of the web, consumers are being exposed to a huge number of choices. On the other side, sellers are challenged by of the diversity of users' interests. In addition to that, nowadays, it is easy to perform a large number of transactions in a small amount of time and as a result, the sale volume is increasingly growing. Recommender Systems (RS) are designed to address the natural dual need by both consumers and sellers. Current RS can be classified as Content Filtering and Collaborative Filtering approaches. RS generally use Collaborative Filtering because it uses the entire user-base information explicitly to recommend items. We will be focusing on Collaborative-Filtering and, for the rest of the paper, we will be referring to it as CF. The process of computing the similarity between users in CF requires that those users should share common rated items which is not practical in real life as systems generally process a  large number of items. As a result, it is very likely to happen that two random users have no common items and the RS fail to predict a rating.\\ 
\indent{Inspired} by the work presented in \cite{Massa}, we propose a novel solution which uses trust network to achieve better rating prediction accuracy and a higher rating coverage. 
Our key contributions are:\\ 
\textnormal{\indent{\textbf{Investigating factors effecting CF technique}}: We perform
extensive experimental evaluation of the CF algorithm performance on three real-world datasets (MovieLens, Epinions, Flixster) and show that CF performs differently on every dataset. 
\\
\indent{\textbf{\emph{Trust-Aware Neighbourhood} Algorithm}}: We propose a novel approach to overcome the expensive execution time of \cite{Massa} by limiting the trusted neighbours involved in the rating prediction. We show a significant increase of the rating coverage and an improvement of the rating accuracy on Epinions and Flixster datasets.
\\
\indent{\textbf{\emph{Hybrid Trust-Aware Neighbourhood} Algorithm}}: Focusing on cold-start users, we propose an algorithm that incorporates user rating behaviour and trusted network to increase the neighbourhood and, as a results, increase the item rating coverage without the detriment to prediction accuracy. }}\vspace{-1.4em} 
\section{Background and Related work}
\setlength{\parindent}{0.3cm} 
\textnormal{\indent{The} basic idea behind CF technique is to recommend items for an active user by finding users who have similar rating behaviour as that active user. 
The rating prediction for an item j by user a $P_{a,j}$ is calculated by 
the following standard formula \cite{Su}:
\begin{equation}
\label{eq:weighted sum of others ratings}
P_{a,j}=\bar{r}_a+\frac{\sum_{u \in U}(r_{u,j}-\bar{r}_u)\cdot w_{a,u}}{\sum_{u \in U}|w_{a,u}|}
\end{equation}
Where {$\bar{r}_a$} and  {$\bar{r}_u$} are the average ratings for user \emph{a} and \emph{u} on all other items and {$w_{a,u}$} is the weight between user \emph{a} and \emph{u}.}\\
\indent{Using} social information to improve RS is a recent active research field: 
In \cite{Liu}, 
different approaches are implemented to select neighbours contributing to CF recommendation, with best rating accuracy achieved when nearest neighbours are aggregated with social friends. Even though authors proposed interesting approaches to incorporate social networks in RS, they used a laboratory environment to build the dataset (not an existing real life dataset).There are many example of trust-based models and SocialMF \cite{flixster} is one of them. \cite{flixster} uses Matrix Factorization technique and trust propagation mechanism for recommendation in social networks where social influence is injected in the recommendation model. Even though this model improved the rating prediction accuracy of the RS, the training phase execution time to learn the parameters of the model was expensive. Additionally, the authors did not consider the RS rating coverage evaluation metric. On the contrary, the main contribution in \cite{Hwang} was to improve the rating coverage by incorporating more users in the recommendation process by considering trustors as well as trustees as neighbours in the recommendation. While coverage was increased, the traditional techniques accuracy was preserved. In our work, we aim to increase the rating accuracy as well as the coverage. Similar results to \cite{Hwang} where achieved in \cite{Victor} where authors combined the best of traditional CF technique and trust-based recommendation techniques to present the \emph{EnsembleTrustCF}. While authors focused on a special type of items (controversial ones) ignoring other points (ex. scalability, execution time), our work covers all items and aims to increase the accuracy and coverage in an efficient manner. In \cite{Ma}, trust and distrust relationships are used separately to improve the recommendation prediction process. Authors suggest Matrix Factorization technique with regulations terms constraining the trust and distrust relationships between users and show a rating accuracy improvement. Even though this model is proven theoretically to be scalable over large datasets, it has only been tested  on one relatively small dataset without studying the execution time of the model.
\vspace{-0.7em} 
\section{Factors Affecting CF Technique}
\textnormal{\indent{Though} there is a huge commercial interest in CF technique, there is little published research on the relative performance of various factors including the similarity computation techniques and the dataset characteristics. User-based algorithm is a type of CF that computes the predicted rating $P_{a,j}$ based on rating information from neighbours defined as the set of users who rated the same item as active user $u_a$. This technique is a predominant type of CF algorithm that shows good prediction accuracy in practice \cite{Su}. One critical step in this algorithm is to compute the similarity between users $w_{a,u}$ which can be performed using various algorithms. In this work, we chose the following four standard similarity computation techniques: "Correlation", 
"Vector-Similarity", 
"Inverse-User-Frequency" 
and "Case-Amplification"  
that are used in \cite{MassaCode}. 
To evaluate every technique, we use the following  evaluation metrics: MAE, RMSE, MAUE (The average value of MAE for all users) and RMSUE (The average value of RMSE for all users)\cite{Su}. We will only be showing results of RMSUE due to space constraints.}
\vspace{-1.4em} 
\subsection{Datasets}
\setlength{\parindent}{0.3cm} 
\textnormal{\indent{We} experimented with three different real-world large datasets. The film recommendation MovieLens dataset, 
 Epinions dataset where products are reviewed by users 
and Flixster dataset that comes from a social networking service where every user has 
a list of friends coming from different social sites (e.g Facebook, Myspace). MovieLens and Epinions's ratings range from 1 to 5 while Flixster's ratings range from 1 to 10.
Table \ref{table:general statistics of datasets} summarizes the global statistics of the three used datasets:}
\vspace{-1.5em} 
\begin{table}[!ht]
\caption{MovieLens, Epinions and Flixster datasets Statistics}
\label{table:general statistics of datasets}
\centering
\small
\begin{tabular}{|l|l|l|l|l|l|}
 \hline
Name         & Users & Items & Ratings   & Spars. & Avg Rating \\ \hline
MovieLens    & 943      & 1,682     & 100,000      & 93.67\%  & 106.04                      \\ \hline
Epinions & 49,290    & 139,738   & 664,824      & 99.99\%  & 13.49                          \\ \hline
Flixster     & 1 million & 49,000    & 8.2 million & 99.98\%     & 55.00                        \\ \hline
\end{tabular}
\end{table}
\vspace{-3em} 
\begin{table}[!ht]
\centering
\small
\caption{Percentage of different "cold-start" user types in the datasets}
\label{table:cold start in Flixster and Epinions}
\begin{tabular}{|l|l|l|l|}
\hline
Users Types                 & \textit{MovieLens} &\textit{Flixster} & \textit{Epinions} \\ \hline
\textbf{"No-rating" }Users  & 0 & 90.19             &  18.51                 \\ \hline
\textbf{"Few-rating"} Users & 0 & 7.81             &  34.31                 \\ \hline
\textbf{Total "Cold-start"} Users & 0 & 98.00             & 52.83                 \\ \hline
\end{tabular}
\end{table}
\vspace{-1.5em} 

\subsection{Similarity Computation Effect on CF}
\textnormal{\indent{Our} CF algorithm implementation is an extension to the "state of the art" RS library available in \cite{MassaCode}.}
\begin{figure}[!ht]
\centering
\small
\includegraphics[width=2.6in] {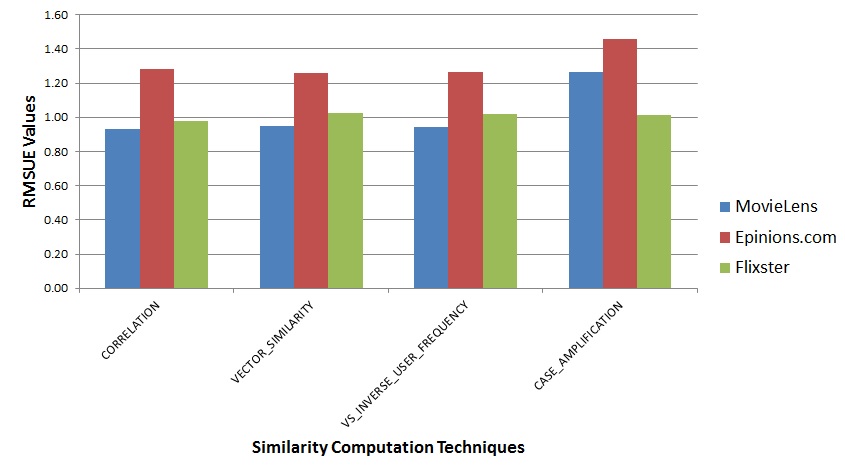}
\caption{Similarity computation effect on MovieLens, Epinions and Flixster datasets}
\label{fig:EpinionsFixsterEpinions}
\end{figure} 
\vspace{-1em}From Fig.\ref{fig:EpinionsFixsterEpinions} we notice that \emph{Case Amplification} generally is the worst among the other algorithms. This is because, unlike the others, this technique does not consider the weight value between users when computing the prediction. Next, MovieLens dataset has the lowest RMSUE value (0.9305 for "Correlation") compared to Epinions and Flixster datasets (1.2844 and 0.9762 respectively). We can further observe that CF algorithm performs better in Flixster dataset compared to Epinions.\\
\indent{From} those observations, our intuition is that (regardless of the similarity computation technique) the CF algorithm performs differently in the three datasets based on its characteristics. From Table \ref{table:cold start in Flixster and Epinions}, MovieLens dataset has 0\% of cold-start users (who have rated at least 20 items) and this may be the main reason behind its low RMSE values, which are statistically significantly better than the Flixster ones at the \emph{p} $<$ 0.01 level. The reason behind the fact that Flixster has better predictions compared to Epinions, even though more than half users are cold-start in both datasets (98.00\% and 52.83\% respectively), may be that every dataset has a different type of cold-start users. Cold-start users may either be users who have no ratings ("No-rating" users) or users with 1-5 ratings ("Few-rating" users). From Table \ref{table:cold start in Flixster and Epinions}, most cold start users in Flixster are "No-Rating" users (90.19\%) and, on the other hand, most cold-start users in Epinions are "Few-Rating" ones (34.31\%). "No-Rating" users do not effect the rating prediction because they do not contribute in the rating computation (they are not considered neighbours). On the contrary, "Few-Rating" users may badly effect the rating prediction because, even though few items are shared between the active user and those users, i.e. the rating behaviour is different, they still contribute in the prediction. Furthermore, the average rating in Flixster is 55.00 items per user but is 13.49 in Epinions (Table \ref{table:general statistics of datasets}). More ratings per user means more users contributing in the CF technique and leads to better rating prediction.
\section{Trust-Aware Neighbourhood}
\subsection{Overview}
\textnormal{\indent{It} is impossible to gather ratings from all users on all items. If we use "Trust network", we can overcome this limitation where trust network can be gathered from external social sites. 
Motivated by this, we propose \emph{Trust-Aware Neighbourhood} (\emph{T-A Neighbourhood}) algorithm where we suggest a new definition of "neighbourhood". In \cite{Massa}, the distance for each user who rated item \emph{i} from user \emph{u} in the trust network needs to be computed. To void this expensive computation that we have tested and it took more than weeks to finish, we instead consider only the trusted users of \emph{u}. Our intuition is that there is a similar rating behaviour between users who trust each other. The method is not compared to baseline method because of computations limitation.
Additionally, we expect the rating coverage to increase, especially with rich trust network datasets. Unlike \cite{Massa} where the weight in Eq.\ref{eq:weighted sum of others ratings} is considered an estimated trust value, the proposed algorithm uses the "Correlation" as the weight which is the key point to overcome the scalability issue in \cite{Massa} where we don't have to compute the minimum distance between the target user and all of its neighbours.\\}
\vspace{-2em}
\begin{figure}[!ht]
\centering
\includegraphics[width=3in]
{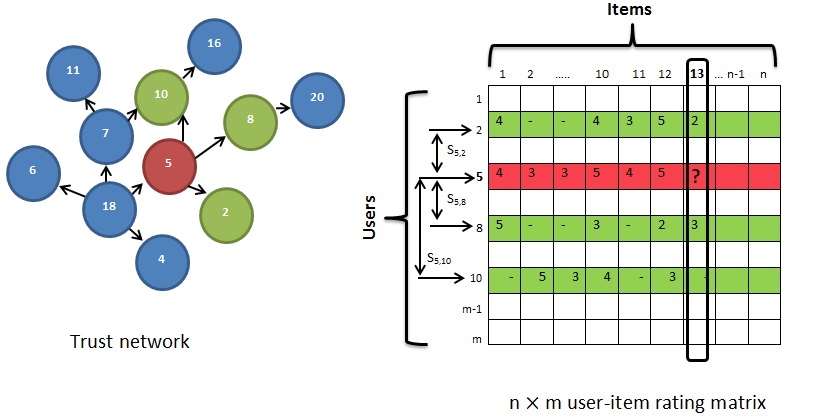}
\caption{\emph{T-A Neighbourhood} Example}
\label{fig:Trust-Aware NeighbourhoodCF}
\end{figure}
\vspace{-1em}\\
\indent{Figure \ref{fig:Trust-Aware NeighbourhoodCF}} presents a demonstrative example of running \emph{T-A Neighbourhood} algorithm on an n$\times$m user-item rating matrix $\mathscr{A}$. For predicting $P_{5,13}$: First, we consider the list of neighbours of $u_5$ which is \{10,8,2\}, as opposed to \{8,2\} using \cite{Massa} technique ($P_{10,13}$ is not available) where d=1 or \{8,2,20,16\} where d=2. After computing the weights between users, Eq.\ref{eq:weighted sum of others ratings} is used to compute $P_{5,13}$. The demonstrative example shows how we can limit the neighbourhood list from \{8,2,20,16\} using \cite{Massa} to \{10,8,2\} using \emph{T-A Neighbourhood} in order to reduce the expensive execution time.
\vspace{-1em}
\subsection{Evaluation}
\textnormal{ \indent{The} datasets used for evaluation are: Epinions which contains 487,181 issued trust statements 
and 
7.2 direct neighbour per user 
and Flixster dataset (users in Flixster can specify a list of friends so, for the context of trust networks, we assume a friend is a trusted user) which has 26.7 million social friendship relation and a 27 average friends per user.
Regarding the rating prediction accuracy (Fig.\ref{fig:Trust-Aware NeighbourhoodCF Results}), we can observe that, generally, \emph{T-A Neighbourhood} CF algorithm provides better rating prediction accuracy compared to traditional CF. For example, in Epinions dataset, the relative error was reduced from 1.2855 to 1.1509 on a random set containing 1,000 users. \emph{T-A Neighbourhood} is statistically significantly better than the traditional CF algorithm for both datasets Epinions and Flixster at the \emph{p} $<$ 0.01 level for all data samples. The difference in the results improvement in Epinions dataset is more obvious than in Flixster. Furthermore, the \emph{T-A Neighbourhood} performance is more substantial in smaller subsets of Epinions dataset where we used only 100, 500 and 1,000 users and we will later discuss the reason behind this behaviour.
\vspace{-1em}
\begin{figure}[!ht]
\centering
\includegraphics[width=3.1in]{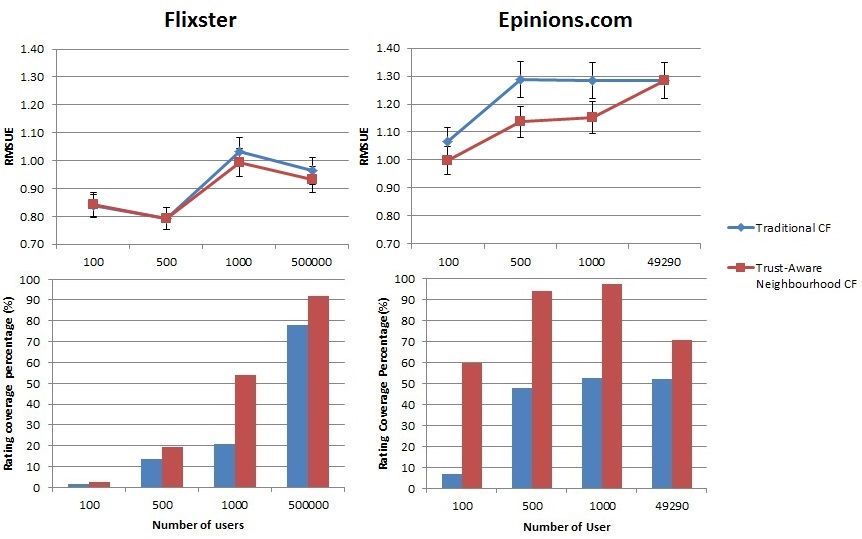}
\caption{\emph{T-A Neighbourhood} RMSUE/Rating Coverage on Flixster \& Epinions datasets for different user sets}
\label{fig:Trust-Aware NeighbourhoodCF Results}
\end{figure}
\vspace{-1em}
Fig. \ref{fig:Trust-Aware NeighbourhoodCF Results} also shows an increase of the rating coverage when using \emph{T-A Neighbourhood} algorithm compared to the traditional CF technique. For instance, in Flixster dataset, the rating coverage is almost doubled from 20.65\% to 53.98\% (1,000 user). Additionally, Flixster's rating coverage is higher than the Epinions's one (especially with 50,000 users).\\}
\textnormal{\indent{Results} showed a rating accuracy increase which may be a result of increasing the rating coverage. Our intuition is that the rating coverage increased in \emph{T-A Neighbourhood} compared to the traditional CF technique for two reasons: First, a user's average number of ratings is low compared to the average number users an active user trust. In Flixster, the average ratings per user is 8.2 (considering the "No Rating" cold-start users) compared to an average of 27.0 direct trusted user. Second, the amount of common items between neighbours in \emph{T-A Neighbourhood} technique is higher than the amount of those items in the traditional CF technique. This my be the main reason behind having an improvement of rating prediction accuracy in Epinions dataset (Fig \ref{fig:Trust-Aware NeighbourhoodCF Results}).  Furthermore, experiments of having small subsets of the whole dataset presents the situation where we have a small item-rating matrix compared to a large trust network. Having a rich trust network does not only increase the number of neighbours for an active user, it also increase the amount of common items and, as a result, achieves a substantial improvement over the traditional CF algorithm. This behaviour is more obvious in Epinions subsets compared to Flixster subsets mainly because the average rating per user in Epinions is low  compared to a higher user rating average in Flixster (13.49 and 55.00 respectively from Table \ref{table:general statistics of datasets}).
Note that running \emph{T-A Neighbourhood} algorithm on cold-starts, with fewer than 5 ratings, produces bad rating coverage (less than 1\% in Epinions dataset). Next, will present a solution to increase the rating coverage for such users.}
\textnormal{\\\indent{The} execution time of \emph{T-A Neighbourhood} takes {$O(m \times n \times t)$} steps in every prediction, where \emph{m} is the number of users, \emph{n} is the number of items and \emph{t} is the number of trust statements. This execution time depends on \emph{m} and \emph{t} values (as \emph{n} is usually small). \emph{T-A Neighbourhood} algorithm will be less expensive than \cite{Massa} if $t \times n$ value is smaller than $m^2$. 
We observed from the experiments a drastic difference in  execution time between the two algorithms where \cite{Massa} took 63 hours to run compared to 20 minutes of execution time of \emph{T-A Neighbourhood} technique. This is mainly because the number of items \emph{n} is small compared to the size of trust network \emph{t} and the product \emph{n}$\times$ \emph{t} is much less than $m^2$.}
\vspace{-1em}
\section{Hybrid Trust-Aware Neigh. }
\subsection{Overview}
\textnormal{\indent{To} overcome the low cold-start item rating coverage of \emph{T-A Neighbourhood} CF algorithm, we propose a new approach called \emph{Hybrid Trust-Aware Neighbourhood} (H T-A Neigh.). This novel technique suggests a new definition for "neighbourhood" suitable only for cold-starters users. According to the "Hybrid" approach, a neighbour is a user who rated an item \emph{i} or is a trusted user by the active user. We expect that the number of neighbours for an active cold-start user to increase and as a result, the rating coverage will increase.\vspace{-1.3em}
\begin{figure}[!ht]
\centering
\includegraphics[width=3in]{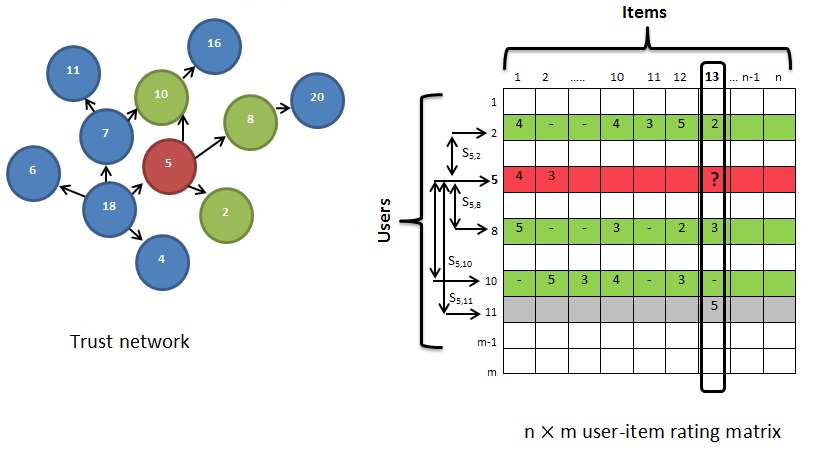}
\caption{\emph{H T-A neighbourhood} Example}
\protect
\label{fig:Hybrid trust-aware neighbourhood cf process}
\end{figure}
\vspace{-1.5em} \\}
\textnormal{\indent{Figure} \ref{fig:Hybrid trust-aware neighbourhood cf process} presents the process of \emph{H T-A Neigh.} algorithm on an \emph{n}$\times$\emph{m} user-item rating matrix. To compute $P_{5,13}$, for example, we first get the directly connected users to $u_5$ which are \{2,8,10\}. Next, we get users who rated $i_{13}$ \{2,8,11\}. The final list of neighbours is  \{2,8,10\} $\cup$ \{2,8,11\} which is \{2,8,10,11\}. The "Correlation" between $u_5$ and any user $\in \{2,8,10,11\}$ is computed to predict $P_{5,13}$ using Eq.\ref{eq:weighted sum of others ratings}.}
\vspace{-2em}
\subsection{Evaluation}
\textnormal{\indent{Experiments} showed that Epinions cold-start user's RMSUE value was reduced from 1.49 with traditional CF to 1.47 with \emph{H T-A} approach and similar results were achieved for Flixster dataset. Furthermore, we achieved the same RMSUE values when using \emph{T-A Neigh.} and the \emph{Hybrid} approach. The rating accuracy was not further improved  with the \emph{Hybrid} approach because few users are added to the recommendation (cold-start users have rated no more than 5 items) and this may not have an noticeable effect on the prediction accuracy.
Experiments also showed a jump in the rating coverage when using \emph{H T-A} over both traditional CF and \emph{T-A Neigh.}. Epinions's rating coverage increased from almost 0\% to 20.57\% while Flixster's coverage was almost doubled (59\% to 98\%). Flixster's item rating coverage was reduced from 59\% using traditional CF to 20\% using \emph{T-A Neighbourhood} algorithm because we used a random set of 5,000 trust statement (not the whole trust network) for memory space limitations. The rating coverage increase comes from the fact that we are incorporating more neighbours so the probability of having common items among users increases which means RS is able to rate more items. Additionally, Flixster dataset reached a rating coverage close to 100\% which means the RS can predict a rating for almost all items due to the high average user rating.}
\vspace{-1em}
\section{Conclusion}
\textnormal{\indent{We} proposed \emph{T-A Neigh.} algorithm that incorporates trust network in the rating prediction process and shows a substantial improvement in the item rating coverage and accuracy especially in Epinions dataset which has fewer ratings per user. Focussing on cold-start users, we proposed \emph{H T-A Neigh.} algorithm which reached a near complete rating coverage 
for Flixster dataset while keeping the same \emph{T-A Neigh.} rating accuracy. We can further augment the work to an elastic cloud based environment implementation which  takes as input a rating prediction to be computed $P_{u,i}$ and a specific budget limit. If the user has a large set of trusted users, then  $P_{u,i}$ will be computed with a small value of the maximum propagation distance. If the user has few trusted users, then the algorithm increases the maximum propagation distance till reaching the budget limit. We  expect this elastic algorithm to achieve high and efficient rating prediction accuracy value computation.}
\vspace{-1em}


\bibliographystyle{abbrv}


\end{document}